\documentclass[superscriptaddress, 
a4paper,twocolumn, showpacs, nofootinbib,amsmath,amssymb,aps,prc,floatfix]{revtex4-1}

\usepackage{graphicx}
\usepackage{dcolumn}
\usepackage{bm}
\usepackage{hyperref}
\usepackage{color}

\begin{document}

\title{Constraining the relativistic mean-field model equations of state with gravitational wave observation }

\author{Rana Nandi}
\affiliation{Department of Nuclear and Atomic Physics, Tata Institute of Fundamental Research, Mumbai 400005, India}

\author{Prasanta Char}
\affiliation{INFN Sezione di Ferrara, Via Saragat 1, I-44100 Ferrara, Italy}

\author{Subrata Pal}
\affiliation{Department of Nuclear and Atomic Physics, Tata Institute of Fundamental Research, Mumbai 400005, India}

\begin{abstract}

The first detection of gravitational waves from the binary neutron star merger
event GW170817 has started to provide important new constraints on the nuclear equation 
of state at high density. The tidal deformability bound of GW170817 combined 
with the observed two solar mass neutron star poses serious challenge to theoretical
formulations of realistic equation of state.
We analyze a fully comprehensive set of relativistic nuclear mean-field theories 
by confronting them with the observational bounds and the measured neutron-skin thickness.
We find that only few models can withstand these bounds which predict a stiff overall
equation of state but with a soft neutron-proton symmetry energy. Two possible 
indications are proposed: Circumstantial evidence of hadron-quark phase transition 
inside the star and new parametrizations that are consistent with ground state
properties of finite nuclei and observational bounds. 
Based on extensive analysis of these sets, an upper limit on the radius of a
$1.4M_\odot$ neutron star of $R_{1.4}\lesssim 12.9$ km is deduced.

\end{abstract}

                             
\maketitle

\section{Introduction}

The equation of state (EOS) of nuclear matter, characterizing the relation between energy 
density and pressure of the system, 
has been the cornerstone in defining the structure of rare isotopes \cite{Brown:2000pd},
collective properties in nucleus-nucleus collisions \cite{Li:2008gp,Tsang:2008fd}, 
and the structure of neutron star \cite{Lattimer:2004pg,Oertel:2016bki}.
Yet the predictions of these observables are largely restricted due to incomplete
knowledge of the EOS. 

While a first principle calculation of finite density quantum 
choromodynamics in lattice gauge theory is plagued by the sign problem 
\cite{deForcrand:2010ys}, sophisticated nuclear many-body theories 
\cite{Dutra:2014qga,Oertel:2016bki,Lattimer:2015nhk} have served as a promising prospect.
These calculations by design reproduce the nuclear matter properties at the saturation density. As a consequence, 
the lower and higher density predictions of these EOSs are very diverse and 
remain largely unconstrained.
Particularly uncertain is the supranuclear density behavior of nuclear symmetry energy
$E_{\rm sym}(\rho)$ and thus the EOS of neutron-rich matter
\cite{Brown:2000pd,Li:2008gp,Lattimer:2004pg}.

The first major observational constraint of the EOS at supra-saturation densities came from
the precise measurements of two massive neutron stars (NS) of masses $(1.928\pm0.017)M_\odot$ 
\cite{Fonseca:2016tux} and $(2.01\pm0.04)M_\odot$ \cite{Antoniadis:2013pzd}.
This would effectively exclude unduly soft EOSs where the matter pressure is 
not sufficient enough to support stars of maximum mass $M_{\rm max} \geq 1.97M_\odot$
against gravitational collapse. Conversely, a stiff EOS with large energy
density and pressure offers an intriguing possibility to produce exotic phases
comprising of hyperons \cite{Maslov:2015msa,Char:2015nea} and quarks \cite{Alford:2004pf}.

The historic detection of gravitational waves (GW) on August 17, 2017 by the
LIGO and Virgo collaborations from the binary neutron star (BNS) merger event GW170817
\cite{TheLIGOScientific:2017qsa} marks the opening of a new possibility to explore the EOS 
at large densities. The GW signal encodes the information of tidal deformation induced
by the strong gravitational field of each star on its companion during the inspiral phase.
The tidal deformability, which depends inherently on the properties of neutron star, 
can be quantified at the leading-order as \cite{Hinderer:2009ca}
\begin{eqnarray}\label{deform}
\Lambda = \frac{2}{3} k_2 \left( \frac{Rc^2}{GM} \right)^5,
\end{eqnarray}
where $k_2$ is the tidal Love number that depends on the EOS.
The large sensitivity of the tidal deformability 
on the star radius is expected to impose severe constraint on the EOS.

The LIGO-Virgo collaborations inferred a bound on $\Lambda_{1.4} \leq 800$ for 
neutron stars of mass $M = 1.4 M_\odot$ from Bayesian analysis of the GW data under 
the assumption that each star may have a different EOS \cite{TheLIGOScientific:2017qsa}.
Since then, different analysis techniques and model studies 
were undertaken in an effort to constrain the radii and/or maximum mass of neutron stars and the associated EOSs 
\cite{Margalit:2017dij,Radice:2017lry,Fattoyev:2017jql,Annala:2017llu,Most:2018hfd,Nandi:2017rhy,Zhang:2018vrx}
by using the reported $\Lambda_{1.4}$ upper bound.
Recently, an improved analysis of this data, using a common EOS 
for both the stars and with more realistic waveform models, provides 
$\Lambda_{1.4} = 190^{+390}_{-120}$ that translates to a stringent bound of 
$\Lambda_{1.4} \leq 580$ at the $90\%$ confidence level \cite{Abbott:2018exr}.

Complementary laboratory measurements of skin thickness of neutron-rich heavy nuclei 
can provide further important checks on the EOS at subsaturation densities
\cite{Horowitz:2000xj,RocaMaza:2011pm,Sharma:2009zzk}.
Remarkably, the neutron-proton asymmetry pressure that determines the skin
in a nucleus of radius $R_{\rm nuc} \sim 10$ fm is essentially the same pressure 
that dictates the radius $R \sim 10$ km of a neutron star \cite{Brown:2000pd}.
The PREX measurement at the Jefferson Laboratory \cite{Abrahamyan:2012gp}
for the neutron-skin thickness of $^{208}$Pb, $R^{208}_{\rm skin} = 0.33^{+0.16}_{-0.18}$ fm,
may well be employed to impose additional constraints. 
However, a definitive data-to-theory comparison would require a substantial reduction 
in the statistical error as planned in the future PREX-II experiment.

The synergy between low and high density physics of nuclear matter can be suitably explored 
using the relativistic mean field (RMF) theory that provides a natural Lorentz 
covariant extrapolation from sub- to supra-saturation densities
\cite{Dutra:2014qga,Lattimer:2015nhk}. The RMF models offer a comprehensive 
framework that successfully describes several finite nuclei
properties and finds large applications in studies of NS structure. 

In this article we have extensively analyzed 269 various EOSs predicted by the RMF models by 
using the latest observational bounds on neutron stars and measured finite nuclei properties.
From the analysis we infer plausible bounds on the radius of neutron stars.
We further show that the recent stringent bound on tidal deformability can be reconciled
with the appearance of quark phase inside the neutron stars. New relativistic parametric sets 
are introduced that simultaneously describe the finite nuclei properties and high-density 
observational constraints.

\section{Set-up}

In the original RMF model \cite{Walecka:1974qa,Boguta:1977xi,Serot:1997xg}, the interaction between the 
nucleons is described via the exchange of scalar-isoscalar $\sigma$ meson and vector-isoscalar $\omega$ meson. 
Over the years, the model has been refined by including other mesons (such as vector-isovector 
$\rho$ meson and scalar-isovector $\delta$) and introducing non-linear self-interaction as well as 
cross-coupling terms for all the mesons
\cite{Horowitz:2000xj,Lalazissis:1996rd,Centelles:1997wy,Typel:1999yq,Liu:2001iz,Horowitz:2002mb,
Bunta:2003fm,ToddRutel:2005fa,Chen:2014mza}.

Based on the form of the interactions in the Lagrangian density, the 269 RMF models 
\cite{Oertel:2016bki,Dutra:2014qga} are broadly recognized as:
NL-type (with nonlinear $\sigma$ term) \cite{Centelles:1997wy,Liu:2001iz},
NL3-type (NL3 and S271 families with additional $\sigma-\rho$ and $\omega-\rho$ couplings)
\cite{Lalazissis:1996rd,Horowitz:2002mb},
FSU-type (FSU and Z271 families with an additional nonlinear $\omega$ coupling) 
\cite{Bunta:2003fm,ToddRutel:2005fa,Chen:2014mza},
BSR-type (BSR and BSR* families with more nonlinear couplings; BSR does not have nonlinear $\omega$ coupling)
\cite{Dhiman:2007ck,Agrawal:2010wg},
and DD (with density-dependent couplings) \cite{Typel:1999yq}.
The associated coupling constants are obtained by sophisticated fitting
procedures to the binding energies and charge radii of finite nuclei and/or to the nuclear
matter properties at the saturation density $\rho_0$. 

The total energy per nucleon, i.e. the EOS, 
$E(\rho,\delta) = E_0(\rho) + E_{\rm sym}(\rho)\delta^2$, is sum of the
symmetric nuclear matter (SNM) energy per nucleon $E_0(\rho)$ and nuclear symmetry energy $E_{\rm sym}(\rho)$, 
where $\delta = (\rho_n-\rho_p)/\rho$ is the isospin asymmetry and $\rho_n$, $\rho_p$ and $\rho$ 
are respectively the neutron, proton and nucleon densities \cite{Li:2008gp,Tsang:2008fd}. 
Large-scale comparison \cite{Oertel:2016bki} of experimental data from finite nuclei and 
heavy-ion collisions with various model calculations
have provided reliable bounds on incompressibility of SNM 
$210 \leq K_\infty = 9\rho_0 |\partial^2 E/\partial\rho^2|_{\rho_0} \leq 280$ MeV, symmetry energy 
$28\leq E_{\rm sym}(\rho_0) \leq 35$ MeV and its slope parameter 
$30 \leq L= 3\rho_0 |\partial E_{\rm sym}(\rho)/\partial\rho|_{\rho_0}\leq 87$ MeV at the
saturation density $\rho_0$. By imposing these current experimental bounds, 
67 RMF models out of 269 sets are found to survive. We will examine the impact of observational
bounds and measured neutron-skin thickness on these EOSs without altering the 
parameters in each model.

\section{Results}

\begin{figure}[t]
\centering
\includegraphics[width=\columnwidth]{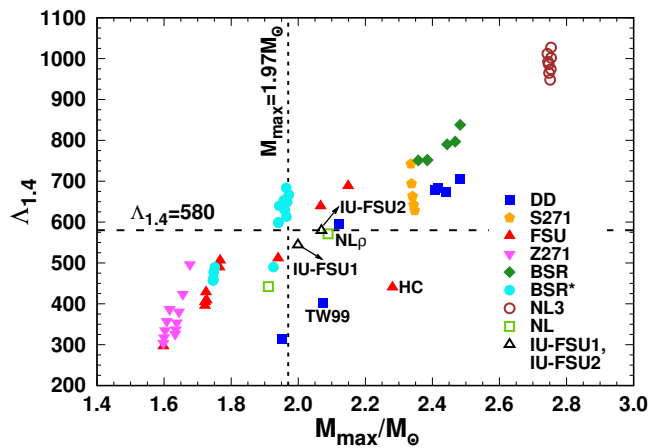}
\caption{Tidal deformability $\Lambda_{1.4}$ of neutron star of mass $1.4M_\odot$ 
versus maximum mass $M_{\rm max}$ for all the RMF EOSs. 
The horizontal and vertical lines respectively refer to the recent upper bound
$\Lambda_{1.4}=580$ of GW170817 data \cite{Abbott:2018exr}
and lower bound $M_{\rm max}=1.97M_\odot$ from the observed pulsar PSR J0348+0432
\cite{Antoniadis:2013pzd}.}
\label{fig:Mmax}
\end{figure}

Figure \ref{fig:Mmax} presents the prediction of tidal deformability
$\Lambda_{1.4}$  (for mass $M=1.4M_\odot$) as a function of maximum mass $M_{\rm max}$ 
of stars for the 67 RMF EOSs. 
Models those do not support stars of $M_{\rm max}=1.97M_\odot$ have essentially soft 
isospin-symmetric nuclear matter EOS $E_0(\rho)$ which largely dictates NS mass at 
high density. In contrast, the deformability $\Lambda \sim R^5$ (hence NS radius)
is sensitive to the density-dependent symmetry energy $E_{\rm sym}(\rho)$ 
at $\rho \sim 2\rho_0$. The tidal deformability constraint
$\Lambda_{1.4} \leq 800$, inferred from the first analysis of GW170817 event
\cite{TheLIGOScientific:2017qsa}, combined with the lower bound on maximum mass
allow sizable number of RMF EOSs to survive, as can be seen from Fig. \ref{fig:Mmax}.
The current tight bound on $\Lambda_{1.4} \leq 580$ \cite{Abbott:2018exr} 
rules out majority of the EOSs and supports only three existing models with 
rather soft $E_{\rm sym}(\rho) \sim 46$ MeV at $\rho\approx 2\rho_0$, namely 
NL$\rho$ \cite{Liu:2001iz} (NL-type EOS with $\sigma$ self-couplings), 
HC \cite{Bunta:2003fm} (FSU-type EOS with nonlinear $\omega,\rho$), and
TW99 \cite{Typel:1999yq} (a density-dependent EOS). We also note that TW99 
set provides a tidal deformability of $\Lambda_{1.4} \approx 400$.

\begin{figure}[t]
\includegraphics[width=\columnwidth]{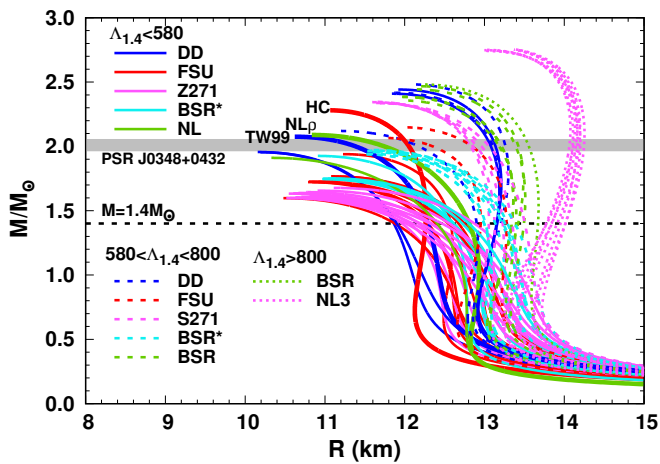}
\caption{Mass-radius relation of neutron stars predicted for all the RMF 
EOSs that fulfill various $\Lambda_{1.4}$ bounds.} 
\label{fig:mr}
\end{figure}

\begin{figure}[ht]
\includegraphics[width=\columnwidth,angle=-90]{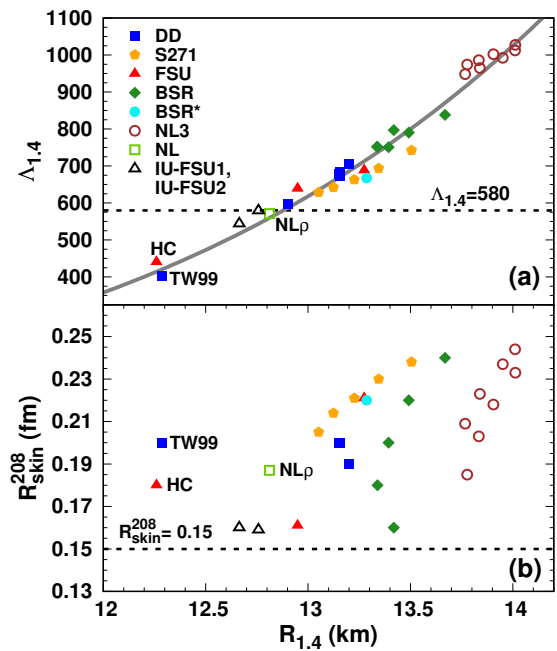}
\caption{(a) Correlation between tidal deformability $\Lambda_{1.4}$ and radius $R_{1.4}$
of neutron stars of mass $M=1.4M_\odot$. The solid line represents the fit
$\Lambda_{1.4} = 1.53\times 10^{-5} (R_{1.4}/{\rm km})^{6.83}$.
(b) Correlation between neutron-skin thickness of $^{208}$Pb nuclei $R_{\rm skin}^{208}$ 
and $R_{1.4}$.
The results are for EOSs that support stars with $M_{\rm max} \geq 1.97M_\odot$.} 
\label{fig:skin}
\end{figure}

To explore the impact of tidal deformability constraint on the entire structure of a
star, we display in Fig. \ref{fig:mr} the mass-radius relation of stars for all the
EOSs that are subjected to various $\Lambda_{1.4}$ bounds. 
The resulting correlation between $\Lambda_{1.4}$ and radius $R_{1.4}$ (for a 
$1.4M_\odot$ star), computed for the EOSs that support maximum mass larger than $1.97M_\odot$, 
is shown in Fig. \ref{fig:skin}(a).
In general, the increase of $R_{1.4}$ with $\Lambda_{1.4}$ has the natural explanation 
that $\Lambda$ quantifies the deviation of gravitational field of a star relative to that 
of a point-mass object \cite{Hinderer:2009ca}.
The exceedingly stiff NL3-type EOSs \cite{Lalazissis:1996rd,Horowitz:2002mb} generate 
stars with large $M_{\rm max} \approx 2.7M_\odot$ but have fairly large 
 $R_{1.4} \approx 13.7$ km (albeit within a narrow range). 
Hence, these stars give $\Lambda_{1.4}>800$ and can be clearly ruled out by the present 
GW data. Interestingly enough, a bound of $580 <\Lambda_{1.4} <800$ suggests 
quite a large variation in the maximum mass
$2.0 \lesssim M_{\rm max}/M_\odot\lesssim 2.5$ but reasonably tight correlation 
between deformability and radii $12.9\lesssim R_{1.4}/{\rm km}\lesssim 13.50$
for these moderately soft EOSs. A plausible stringent 
LIGO-Virgo bound $400 \leq \Lambda_{1.4} \leq 580$ favors EOSs that posses much 
softer $E_{\rm sym}(\rho)$ at density $\rho \sim 2\rho_0$. However, due to super-soft 
total pressure at high densities, most of these EOSs are excluded by the 
$M_{\rm max} \geq 1.97M_\odot$ constraint. As also seen in Fig. \ref{fig:Mmax}, 
only three EOSs:  
NL$\rho$ \cite{Liu:2001iz}, HC \cite{Bunta:2003fm} and TW99 \cite{Typel:1999yq}, 
are just stiff enough to qualify the combined observational and maximum mass constraints.

One important upshot of $M-R$ relation of Fig. \ref{fig:mr} is the large spread 
of radius $R_{1.4}$ when all the stars are considered irrespective of their
$M_{\rm max}$. The maximum mass bound [as shown in Fig \ref{fig:skin}(a)]
enforces a tight correlation of the form 
$\Lambda_{1.4} = 1.53\times 10^{-5} (R_{1.4}/{\rm km})^{6.83}$ which suggests the possibility 
to constrain the radius \cite{Fattoyev:2017jql,Annala:2017llu}
and perhaps the $E_{\rm sym}(\rho)$.
Thus the bound $\Lambda_{1.4} \leq 800$, estimated from the first analysis
of GW170817 \cite{TheLIGOScientific:2017qsa}, translates to $R_{1.4} \leq 13.49$ km,
and the recent stringent constraint $\Lambda_{1.4} \leq 580$
\cite{Abbott:2018exr} provides a strict upper limit of $R_{1.4} \leq 12.87$ km. 
Interestingly, the tidal deformability in NL$\rho$ \cite{Centelles:1997wy} 
is close to the inferred current upper bound and predicts $R_{1.4}\lesssim 12.81$ km.
Albeit, the radius of a NS is known to receive considerable contribution
from the low density crustal equation of state.

It may be mentioned that all the 67 RMF EOSs  are found to be consistent with 
the pressure bound at twice the saturation density of $P(2\rho_0)=3.5_{-1.7}^{+2.7}\times10^{34}$ dyn/cm$^2$ 
(at the $90\%$ confidence level) as extracted from GW170817 data \cite{Abbott:2018exr}.  
Hence, this bound is not very useful to constrain the EOS. In contrast, the bound 
$P(6\rho_0)=9.0_{-2.6}^{+7.9}\times10^{35}$ dyn/cm$^2$ at
$\rho=6\rho_0$ rules out overly soft RMF EOSs. However, this
estimated bound is more than the central pressures 
of the binary components of GW170817 event \cite{Abbott:2018exr} and therefore should be
used with caution.

Complementary and crucial information on $E_{\rm sym}(\rho)$, (i.e the EOS) at subsaturation
densities can be obtained from analysis of skin thickness $R_{\rm skin} = R_n-R_p$ of nuclei,
defined as the difference between the rms radii of neutrons and protons
\cite{RocaMaza:2011pm,Sharma:2009zzk}.
Figure \ref{fig:skin}(b) shows correlation between neutron-skin thickness 
$R_{\rm skin}^{208}$ of heavy $^{208}$Pb and the stellar radius $R_{1.4}$. 
A stiff $E_{\rm sym}(\rho)$ (large slope $L$) induces large values for both 
the skin and star radius. Although the $R_{\rm skin} - R_{1.4}$ correlation is
strong within the same family of EOS \cite{Fattoyev:2017jql},
the spread is quite large when all the EOSs from RMF theory are included.
This relates to the fact, that apart from the slope $L$, the SNM compressibility 
$K_\infty$ also contribute to the $R_{\rm skin}$ and NS radius \cite{Fortin:2016hny}.
This also suggests that constraints on symmetry energy and its slope $L$
from measurements of neutron-skin and tidal deformability would be model-dependent. 

The large statistical uncertainty in the current PREX measurement: 
$R_{\rm skin }^{208}=0.33^{+0.16}_{-0.18}$ fm \cite{Abrahamyan:2012gp}, however,
prevents any definite constraint on the EOSs. For reference, we note that 
while all the parameter sets that predict $R_{\rm skin}^{208} \sim 0.20-0.25$ fm 
are excluded by the observational $\Lambda_{1.4} < 580$ bound, 
the three EOSs \cite{Liu:2001iz,Bunta:2003fm,Typel:1999yq} allowed by this bound
have $R_{\rm skin}^{208} = 0.18-0.20$ fm.
Should future PREX-II experiment confirms central value of skin thickness 
$R_{\rm skin}^{208} > 0.20$ fm with a significantly small statistical error as envisioned, 
then the observationally constrained EOSs: NL$\rho$, TW99, and HC, would be excluded.

\begin{figure}
\centering
\includegraphics[width=\columnwidth,angle=-90]{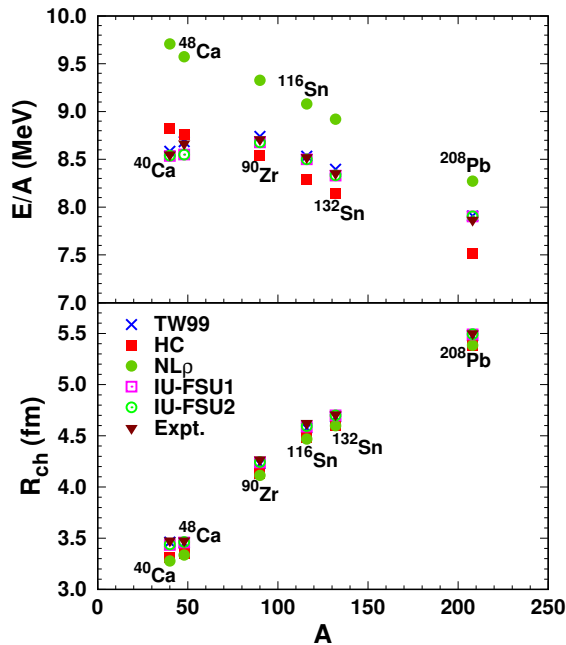}
\caption{Binding energy and charge radius of nuclei calculated for
viable RMF sets and compared with data; see text for details.}
\label{fig:nuclei}
\end{figure}

Any parametric EOS, designed to reproduce nuclear matter properties, 
should also give good description of finite nuclei properties. 
To ascertain this, we have calculated the binding energies and charge radii 
of some light and heavy nuclei for the three parameter sets that satisfy both 
the observational bounds.  The TW99 set which was obtained by including the saturation 
properties of nuclear matter as well as binding energies of some
finite nuclei in the fitting protocol obviously has the best agreement
as seen in  Fig. \ref{fig:nuclei}. In contrast, the other two sets (HC and NL$\rho$) 
which have been fitted to only the nuclear matter saturation properties, fail 
to provide reasonable description of finite nuclei properties.

\section{Implications}

Various parametrizations of RMF model have been generated in the last five decades
that are consistent with nuclear and neutron star properties. 
The tension of RMF models with the current observational data poses intriguing questions: 
Are the GW data an evidence of exotic phases such as quarks inside the NS? 
Is there still scope to device new parameter sets by accommodating all the constraints? 
We will next explore these interesting possibilities.

\begin{figure}
 \centering
 \includegraphics[width=\columnwidth]{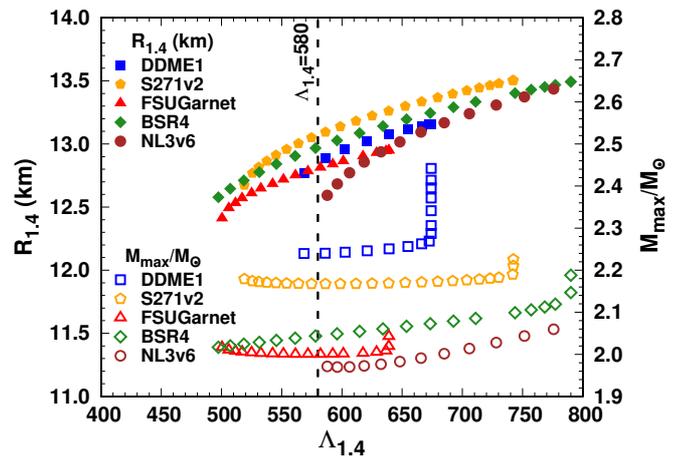}
 \caption{Correlations between $R_{1.4}-\Lambda_{1.4}$ (left scale)
 and $M_{\rm max}-\Lambda_{1.4}$ (right scale) for EOSs with hadron-quark phase transition
 constructed from RMF model for the hadronic phase and bag model for the quark phase.
 In each individual RMF set shown, the bag parameter is varied \cite{Nandi:2017rhy} 
 to generate different EOSs with first-order phase transition using Gibbs conditions.}
 \label{fig:R14vsL14}
\end{figure}

The gravitational waves from the merger of binary neutron stars have the 
potential to investigate the possible existence of deconfined quark phase at high densities
\cite{Radice:2016rys,Most:2018hfd,Most:2018eaw,Bauswein:2018bma}.
The appearance of quarks (or any new degrees of freedom) inside the star 
at $\rho > \rho_0$ softens the EOS resulting in decrease of $M_{\rm max}$ and radius. 
Thus a hadronic EOS which produces $M_{\rm max}\gtrsim 2M_\odot$ for a neutron star could
be a possible candidate for the inclusion of exotic phases.
Such hadronic EOSs can be identified by inspection of Fig. \ref{fig:mr}.

A phase transition from hadron to quark matter in the NS interior, consistent 
with the earlier $\Lambda_{1.4}\leq 800$ constraint,
was recently shown to prevail \cite{Nandi:2017rhy} for realistic parameters in the bag model 
that provides phenomenological description of the quark phase \cite{Bandyopadhyay:1997kh}.
Following the methodology described in Ref. \cite{Nandi:2017rhy}, we generate EOSs with phase
transition by considering one representative 
hadronic EOS from each family of RMF models that gives $M_{\rm max} \geq 1.97M_\odot$ and 
by continuously varying the bag pressure in the range $B_{\rm eff}^{1/4}\simeq 145-200$ MeV. 
A small value of $B_{\rm eff}^{1/4}$ causes early appearance of the quark phase resulting in
small $M_{\rm max}$ and $R_{1.4}$. 
Figure \ref{fig:R14vsL14} illustrates the $\Lambda_{1.4}-R_{1.4}$ and $\Lambda_{1.4}- M_{\rm max}$
correlations obtained from these EOSs with hadron-quark phase transition.
Remarkably, for all these hadronic EOSs (except for NL3v6) we find a range of bag pressures which 
are consistent with the stringent bound of $\Lambda_{1.4}\leq 580$ and  
maximum mass constraint $-$ a possible indication of quark-hadron phase transition in the 
neutron star core. Moreover, these EOSs predict a radius of $R_{1.4} \lesssim 12.94$ km 
close to that found from pure hadronic EOSs.

Finally, we demonstrate how one can generate new EOSs consistent with both the observational 
and experimental data.
Let us consider the original IU-FSU parameter set \cite{Fattoyev:2010mx} which
provides good description of finite nuclei and nuclear saturation properties. We recall that the model 
predicts $\Lambda_{1.4}\simeq 512$, well within the GW170817 bound and $M_{\rm max}=1.94M_\odot$ 
slightly below the $M_{\rm max}$ constraint. The nonlinear self-coupling term
for $\omega$-meson, with coupling constant $\zeta=0.03$, mainly determines the stiffness 
of EOS at high densities \cite{Fattoyev:2010mx}. By fine-tuning $\zeta$ to 0.025 and 0.020,
for example, and refitting other parameters to reproduce the nuclear properties at $\rho_0$,
we construct two new parameter sets: dubbed as IU-FSU1 and IU-FSU2.
Both these sets now generate 
$M_{\rm max}>1.97M_\odot$ and $\Lambda_{1.4}<580$. The resulting correlations involving
$\Lambda_{1.4}$ with $M_{\rm max}$ and $R_{1.4}$ are displayed in
Figs. \ref{fig:Mmax}, \ref{fig:skin}. As expected, these new sets provide 
reasonable description of finite nuclear properties as shown in Fig. \ref{fig:nuclei}.
Interestingly, the NS radii for IU-FSU1 and NL$\rho$ nearly match which may suggest
that $\Lambda_{1.4}\leq 580$ bound translates to $R_{1.4} \lesssim 12.81$ km.

\section{Conclusions}

We have employed observational data from gravitational wave event GW170817 and neutron star 
mass $M_{\rm max}\geq 1.97M_\odot$ in conjunction with laboratory measurements of neutron 
skin thickness to constrain the EOSs within RMF theory. The maximum mass bound excludes several EOSs 
that predict diverse values of NS radius and provides a tight correlation between $R_{1.4}$ and 
$\Lambda_{1.4}$. Whereas, the first inferred bound $\Lambda_{1.4} \leq 800$ translates to a NS radius 
with an upper limit $R_{1.4} < 13.50$ km, the recent improved bound $\Lambda_{1.4} \leq 580$ provides 
$R_{1.4} < 12.88$ km. The strict bound on $\Lambda_{1.4}$ rules out all EOSs, but a few with 
soft $E_{\rm sym}$ at density $\rho \approx 2\rho_0$. If stars have hadron to quark phase transition, 
several EOSs are shown to be consistent with all the measured bounds. Complementary precise estimate of skin
thickness of nuclei that is sensitive to slope of $E_{\rm sym}$ should provide further important checks.

It may be noted that though the phenomenological RMF approach provides a reasonable description of 
the EOS over a wide density range, it does not incorporate the realistic microscopic many-body nuclear 
interactions \cite{Hebeler:2010jx,Rrapaj:2015zba}.
Moreover, the RMF models do not contain the essential features of strong interaction described by
QCD such as chiral symmetry and broken scale invariance \cite{Hanauske:1999ga} at finite
nuclear matter densities. It will be interesting to compare our predictions obtained within
the RMF models with that in the microscopic models.

\end{document}